  \documentclass[aps,pre,twocolumn,groupedaddress,showpacs,nofootinbib]{revtex4-1}

  \usepackage{graphicx}
  \usepackage{amsmath} 
  \usepackage{amssymb}
  \usepackage{placeins}
  \usepackage[T1]{fontenc}
  \usepackage[latin1]{inputenc}
  \usepackage[dvips]{color}
  \usepackage{longtable}

\begin{document}

  \newcommand{\AP}{Ann. Phys.\; }
  \newcommand{\AVN}{Allgemeine Vermessungs - Nachrichten\; }
  \newcommand{\C}{Computing\; }
  \newcommand{\CJP}{Czech. J. Phys.\; }
  \newcommand{\CMP}{Commun. Math. Phys.}
  \newcommand{\CPC}{Computer Physics Communications\; }
  \newcommand{\CS}{Computational Statistics\; }
  \newcommand{\CSQ}{Computational Statistics Quarterly\; }
  \newcommand{\EL}{Europhys. Lett.\; }
  \newcommand{\IJMPB}{Int. J. Mod. Phys. B\; }
  \newcommand{\IJMPC}{Int. J. Mod. Phys. C\; }
  \newcommand{\JAP}{J. Appl. Phys.\; }
  \newcommand{\JCP}{J. Chem. Phys.\; }
  \newcommand{\JLTP}{J. Low Temp. Phys.\; }
  \newcommand{\JMMM}{J. Magn. Magn. Mater.\; }
  \newcommand{\JMP}{J. Math. Phys.\; }
  \newcommand{\JP}{J. Physique\; }
  \newcommand{\JPA}{J. Phys. A\; }
  \newcommand{\JPB}{J. Phys. B\; }
  \newcommand{\JPC}{J. Phys. C\; }
  \newcommand{\JPCh}{J. Phys. Chem.\; }
  \newcommand{\JPL}{J. Physique Lett.\; }
  \newcommand{\JPCS}{J. Phys. Chem. Solids\; }
  \newcommand{\JPSJ}{J. Phys. Soc. Jap.\; }
  \newcommand{\JSP}{J. Stat. Phys.\; }
  \newcommand{\MC}{Math. Comp.\; }
  \newcommand{\NPB}{Nucl. Phys. B\; }
  \newcommand{\Ph}{Physica\; }
  \newcommand{\PF}{Miniworkshop on pattern formation and
                 spatio-temporal chaos, ICTP, Trieste 1997\; }
  \newcommand{\PA}{Physica A\; }
  \newcommand{\PD}{Physica D\; }
  \newcommand{\PL}{Phys. Lett.\; }
  \newcommand{\PLA}{Phys. Lett. A\; }
  \newcommand{\PR}{Phys. Rev.\; }
  \newcommand{\PRp}{Phys. Rep.\; }
  \newcommand{\PRA}{Phys. Rev. A\; }
  \newcommand{\PRB}{Phys. Rev. B\; }
  \newcommand{\PRC}{Phys. Rev. C\; }
  \newcommand{\PRD}{Phys. Rev. D\; }
  \newcommand{\PRE}{Phys. Rev. E\; }
  \newcommand{\PRL}{Phys. Rev. Lett.\; }
  \newcommand{\PS}{Physica Scripta\; }
  \newcommand{\PTP}{Prog. Theor. Phys.\; }
  \newcommand{\RMP}{Rev. Mod. Phys.\; }   
  \newcommand{\RPP}{Rep. Prog. Phys.\; }
  \newcommand{\SPA}{Stochastic Process. Appl.\; }
  \newcommand{\SuS}{Surf. Sci.\; }   
  \newcommand{\SSC}{Solid State Communications\; }
  \newcommand{\ZP}{Z. Phys.\; }
  \newcommand{\ZPB}{Z. Phys. B\; }

  \newcommand{\be}{\begin{equation}}
  \newcommand{\ee}{\end{equation}}
  \newcommand{\bea}{\begin{eqnarray}}
  \newcommand{\eea}{\end{eqnarray}}
  \newcommand{\bean}{\begin{eqnarray*}}
  \newcommand{\eean}{\end{eqnarray*}}
  \newcommand{\ben}{\begin{eqnarray*}}
  \newcommand{\een}{\end{eqnarray*}}
  \newcommand{\bt}{\begin{tabbing}}
  \newcommand{\et}{\end{tabbing}}

 \newcommand{\cF}{{\mathcal F}}
 \newcommand{\cG}{{\mathcal G}}
 \newcommand{\cH}{{\mathcal H}}
 \newcommand{\cK}{{\mathcal K}}
 \newcommand{\cO}{{\mathcal O}}
 \newcommand{\cN}{{\mathcal N}}
 \newcommand{\cq}{complex-$q$ }
 \newcommand{\cS}{complex-$S$ }
 \newcommand{\etl}{{\it et al.}}   
 \newcommand{\mcH}{{\mathcal H}}
 \newcommand{\mcN}{{\mathcal N}}
 \newcommand{\mcO}{{\mathcal O}}
 \newcommand{\mcS}{{\mathcal S}}
 \newcommand{\mbR}{{\mathbb R}}
 \newcommand{\mbT}{{\mathbb T}}
 \newcommand{\req}{real-$q$ }
 \newcommand{\vfi}{\varphi}
 \newcommand{\cZ}{{\mathcal Z}}


\title{Yang-Lee zeros and the critical behavior of the infinite-range two- and three-state Potts models}


  \author{Zvonko Glumac}
  \email[]{zglumac@fizika.unios.hr}
\affiliation{Department of Physics, Josip Juraj Strossmayer University of Osijek, 
             P.O.B. 125,  Trg Ljudevita Gaja 6, 31 000 Osijek, Croatia}

  \author{Katarina Uzelac}
  \email[]{katarina@vrabac.ifs.hr}
\affiliation{Institute of Physics, P.O.B. 304,
             Bijeni\v{c}ka 46, HR-10000 Zagreb, Croatia}


\date{\today}

\begin{abstract}
  The phase diagram of the two- and three-state Potts model with infinite-range interactions, in the external
  field is analyzed  by studying the partition function zeros in the complex field plane. \\
  The tricritical point of the three-state model is observed as the approach of the zeros to the real axis 
  at the nonzero field value. Different regimes, involving several first- and second-order transitions of the 
  complicated phase diagram of the three state model are identified from the scaling properties of the zeros 
  closest to the real axis. The critical exponents related to the tricritical point and the Yang-Lee edge
  singularity are well reproduced. Calculations are extended to the negative fields, where the exact implicit 
  expression for the transition line is derived. 
\end{abstract}


\maketitle

\section{Introduction\label{sec-Intro}}

  Since the famous work of Yang and Lee \cite{YL52,LY52}, the zeros of the partition function in a complex activity plane 
  were extensively studied as a useful tool in the investigation of various aspects of phase transitions, up to giving rise 
  to a new type of criticality related to complex parameters such as the Yang-Lee edge singularity \cite{F78}. 
  Although originally introduced in context of Ising model for explaining singularities related to the second-order phase transition,
  zeros appeare to be useful for deriving the related critical exponents and even distinguishing the nature of a phase transition for 
  a variety of models (see e.g. \cite{K79}, or \cite{BDL05} for a recent review).

  In the case of the Ising model, where all the zeros lie along the circle in the complex plane, the correspondence between the density
  of zeros and thermodynamic potentials is simple and straightforward.
  For the Potts model \cite{P52} with an arbitrary number of states, the layout of Yang-Lee zeros is much more complicated 
  \cite{KL94,KC98,BBCKK00,KC00,BBCKK04} while even more intensive investigations were conducted on  
  Fisher zeros \cite{F65,OKSK68} in this model \cite{K02} and the zeros in the complex-$q$ plane
  \cite{S00,CS00,KC01,GU02}. 
  As far as the Yang-Lee zeros are concerned, 
  it was shown analytically in one dimension \cite{GU94}, and by numerical calculations on small systems in higher dimensions 
  \cite{KC98,KC00}, that, for general $q$, these zeros do not lie on the unit circle of the complex activity plane, and that multiple
  lines, or bifurcations may occur \cite{BBCKK00}.  This is related to a more complicated phase diagram of this model.
  When the number of states of the Potts model is larger than 2, the presence of real, positive external field $h$ does not  necessarily
  destroy the phase transition, but produces a line of first-order phase transitions ending with the tricritical point.
  The model is not symmetric under exchange $h \leftrightarrow - h$, and in the presence of a negative field, a different and less 
  explored \cite{BD2010} transition takes place. 
  All this leads to a more complicated phase diagram in the temperature-field plane, that involves both continuous and  first-order
  phase transitions, and the tricritical points.
  In the complex activity plane, this is reflected  by the fact that zeros do not lie on the unit circle, while more than one line of
  zeros that approach the real axis may appear. 

  The main purpose of this paper is to present, in the special case of the three-state Potts model with infinite-range interactions, 
  an integrated view covering different critical regimes of the entire phase diagram in terms of the behavior of the partition function 
  zeros in the complex activity plane. We also report the results for the Yang-Lee edge singularity above the tricritical temperature
  of this model.

  The infinite-range Potts model is equivalent to the mean-field (MF) approximation of the standard Potts model with short-range (SR)
  interactions \cite{KMS54,W82}, 
  which describes the critical beahaviour of the SR model correctly only in sufficiently high dimensions, above the upper 
  critical dimension. Nevertheless, 
  its phase diagram displays similar diversity to that of the SR model, 
  while it may be approached analytically, using the saddle-point approximation.
  In our approach we apply twice the Hubbard-Stratonovich transformation to obtain the exact two-parameter analytic expression for the 
  free energy of the infinite-range model. The same approach permits us to obtain implicit analytic expressions for the partition function 
  zeros and to calculate zeros close to the real axis in the finite system. 
  The finite-size scaling analysis of a few zeros that lie closest to the real axis then allows us to identify different critical 
  regimes, and calculate the related critical exponents.

  The outline of the paper is following.
  In Sec. \ref{sec-q2} we illustrate the approach for the simple $q=2$ case, which is also used to obtain some estimates about 
  the precision of numerical calculations of the partition function zeros. 
  The same methodology is then applied to the $q = 3$ model in Sec. \ref{sec-q3}.
  We first derive an exact two-parameter expression for the free-energy density to be solved by a saddle-point method, 
  and we discuss its shapes in the two-parameter space, related to different regimes of the phase diagram.
  In Secs. \ref{sec-KMStransitions} and \ref{sec-IsingTransitions}, we derive analytical expressions for the scaling properties 
  of the lowest zeros in different critical regimes, for the cases with $h\geq0$ and $h<0$ respectively. 
  In Sec. \ref{sec-q3Numeric} we present the numerical results for the partition function zeros, and the numerical derivation
  of scaling exponents. Section \ref{sec-zklj} contains the concluding remarks.

  \section{$q = 2$ (Ising) model} \label{sec-q2}

  The energy of the $q$-state MF Potts model with $N$ particles in the external field $H$ has the form
  \be 
  E = -\;\frac{J_0}{N}\;\sum_{i = 1}^{N-1}\;\sum_{j = i + 1}^{N}\;\delta\, (s_i, s_j ) \;
      - \;\,H \;\sum_{i = 1}^N\;\delta\, (s_i, 1 ), 
  \ee
  where $s_i$ denotes the particle at site $i$, which can occupy one of the $q$ Potts states, while
  any two particles interact with the same two-particle interaction $J_0/N$, regardless of their distance.

  The reduced energy may be written in terms of numbers of particles belonging to each of the $q$ states, 
  which for $q = 2$ are denoted by $N_1$ and $N_2$
  \be
  -\,\frac{E}{k_B\;T}  =  \frac{K}{N}\;\left[ \binom{N_1}{2} +  \binom{N_2}{2}   \right]  + h \; N_1, 
  \ee 
  where
  \be 
  K \equiv \frac{J_0}{k_B\,T}, \hspace{2cm}  h  \equiv \frac{H}{k_B\,T} = h_1 + \imath \, h_2, 
  \ee 
  and, by conservation of the total number of particles, $N = N_1 + N_2$. The partition function  
  \be \label{eq:pfq2}
  \cZ_N = \sum_{N_1 = 0}^{N}\;\binom{N}{N_1}\;e^{- E / (k_B\,T)}, 
  \ee  
  after applying the Hubbard-Stratonovich transformation
  \be \label{eq:Gauss}
  e^{A^2} = \frac{1}{\sqrt{2\,\pi}}\;\int_{- \infty}^{+ \infty}\;e^{-\frac{1}{2}\, y^2 + A\, y\; \sqrt{2}}\;d\,y, 
  \ee 
  with
  \be
  A = \sqrt{K\,N} \; \left[\,\frac{N_1}{N} + \frac{1}{2}\;\left(\frac{h}{K}-1 \right)\,\right], 
  \ee
  turns into an integral
  \be\label{eq:pfq2g}
  \cZ_N \; \sim \; \int\; e^{\,- N\, f (S)}\; d S, 
  \ee
  \be
  f (S) = \frac{K \, S}{4} \; ( S + 2) - \ln \Big[1+e^{K S+h} \Big].
  \ee
  It may be solved by a standard saddle-point approximation, which we recall here briefly.
  The corresponding extremal condition may be expressed by 
  \be\label{eq:q2Extrem}
   S = \tanh \frac{K S + h}{2} , 
  \ee
  where $S = (2\,N_1/N) - 1$ is the order parameter,   
  and the model has a second-order phase transition for $h = 0$ and  $K_0 = 2$. \\
  The density of partition function zeros, $g$, which all are located on the imaginary axis of the complex field plane,
  is related \cite{CK97,JK02a,JK02b} to the jump in the order parameter $S$,
  \be \label{eq:DeltaSg}
  S = 2\, \pi g, 
  \ee
  where $S$ is the solution of Eq. (\ref{eq:q2Extrem}).
  Above the critical temperature ($K < 2$) the gap free of zeros opens around the real axis. 
  At its edge  $h_{2, edge}$, which depends on temperature as \cite{K79}
  \be   \label{eq:hedgeIsing}
  h_{2, edge} = \pm\;\left[ \;  2\,\arctan\sqrt{\frac{2}{K}-1} - K\; \sqrt{\frac{2}{K}-1}  \;  \;  \right], 
  \ee
  another second-order phase transition in this model takes place, known as the Yang-Lee edge singularity.

  The exact expressions for the critical exponents are known \cite{K79,W82} for both transitions.
  We shall be interested in scaling properties of the field $h$ and the density of zeros $g$ for which the 
  exact values of their anomalous dimensions in different critical regimes are summarized in Table \ref{tb:yhygq2}.
  \begin{table}   
  \caption{Exact values of anomalous dimensions $y_h$ and $y_g$ as a function of $K$.}\label{tb:yhygq2}
  \begin{ruledtabular}
  \begin{tabular}{ccccccc}
        & & $K < 2$        & & $K = 2$        & & $K > 2$    \\   \hline 
  $y_h$ & & $2/3$          & & $3/4$          & & $1$  \\   
  $y_g$ & & $1/3$          & & $1/4$          & & $1$  \\ 
  \end{tabular}
  \end{ruledtabular}
  \end{table} 
  Note that the anomalous dimensions $y_h$ and $y_g$ are not defined here in the standard way. 
  Due to the infinite range of interactions the space scale has no meaning, so that the size scaling is 
  carried out by scaling the number of particles instead. To to obtain the standard anomalous dimensions for this
  model as obtained when applying the mean-field approximation to the short-range version of this
  model, one should multiply \cite{GU02} the above exponents by $d_c$, the upper critical dimension of 
  the corresponding phase transition (which is equal to 4 for the Ising transition and 6 for the Yang-Lee
  edge singularity). 

  The anomalous dimension of the field is denoted by $y_h$, which may be obtained from the scaling  with size 
  of the distance (in the parameter space) of the imaginary component of the field $h_2$ from its edge value, $h_{edge}$ 
 \be \label{eq:defyh}
  h_2 - h_{edge} \sim  N^{-y_h}.
  \ee
  The exponent $y_g = y_h\,\sigma$ denotes here the anomalous dimension of the density of zeros, the power-law decay of which 
 is described  near criticality by the exponent $\sigma$
  \be \label{eq:defyg}
  g - g_{edge} \sim (h_2 - h_{edge})^{\sigma} \sim \Big( N^{-y_h} \Big)^{\sigma} \sim N^{-y_g}.
  \ee
  Let us now sketch briefly the numerical derivation of these scaling properties by using the partition function zeros 
  closest to the real axis.

  \subsection{Partition function zeros - Numerical results} \label{sec-q2Numeric}

\begin{figure}
  \includegraphics[width=0.9\linewidth]{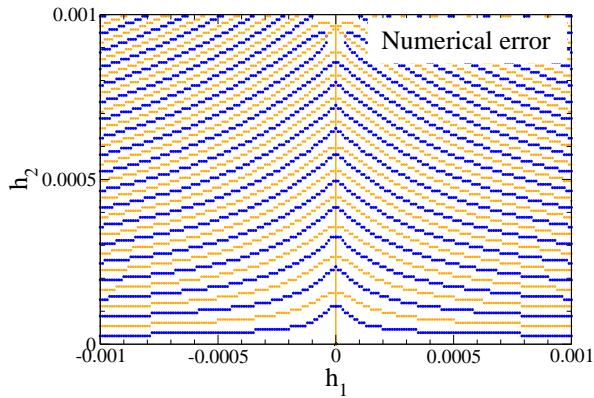}
  \caption{(Color online) Solutions of the equations  $Re \, \cZ_N = 0$ [blue (dark gray) lines] and $Im \, \cZ_N = 0$ 
           [orange (light gray) lines, including the $h_1 = 0$ axis] for $N = 1\,000\,000$ particles, at $K = 2$. 
           For distant points errors can be observed as deviations of intersections from the imaginary axis.
          \label{fg:YLMF-q2N1000001K2t000}    }
  \end{figure}

  For finite $N$, the partition function (\ref{eq:pfq2}), is the polynomial of order $N$ in the complex variable $z = e^h$.
  The usual root-finding routines soon become inapplicable for finding the roots of $\cZ_N$,  as $N$ increases.
  Instead, we look for the numerical solutions of the two equations
  \be
  Re \, \cZ_N  =  0, \hspace{1cm} Im \, \cZ_N  =  0, 
  \ee
  in a complex $(h_1, h_2)$ plane. A typical example of layout of these zeros, given by the intersections of
  dark gray lines and light gray lines $h_1 = 0$, is presented in Fig. \ref{fg:YLMF-q2N1000001K2t000}.  

  Consistent with the Yang-Lee theorem, these zeros lie on the imaginary axis.
  As can be observed in Fig. \ref{fg:YLMF-q2N1000001K2t000}, the 
  precision of such a calculation deteriorates for the remote zeros, which start to deviate from the imaginary axis,
  \cite{KL94}, as a consequence of the accumulation of numerical  errors,  which increase with $N$ and with 
  decreasing $K$.
  However, we do not have to deal with this problem here, since only a few zeros closest to the real axis suffice to 
  detect different regimes and to analyze the corresponding singularities. Our analysis will thus be focused 
  on these zeros, and include a comparison to the exact results allowing the estimation of related errors.

  We calculated the position of the two zeros closest to the real axis for a sequence of different sizes,
  ranging from
  $10^{\,2}$ up to $10^{\,6}$ particles. 
  The distance of the zero closest to the real axis is used
  as an estimate of the position of the Yang-Lee edge at a given temperature, while its convergence to its  value in the thermodynamic 
  limit is expected, according to scaling property (\ref{eq:defyh}), to give the anomalous dimension $y_h$
   \be  \label{eq:h2Conv}
   h_2 (N) = h_{2, \text{extr}} + \frac{\text{const}}{N^{y_2}} + \cdots .
  \ee
  The results of extrapolating data by a simple least-squares fit are presented in Table \ref{tb:q2hedgeExtrImh}.

  The errors of the extrapolated values in Table \ref{tb:q2hedgeExtrImh}
  (as well as in all the following tables) have two origins. 
  First is the pure numerical error, which is relatively small (e.g., for $y_2$  it ranges from
  $10^{-4}$ to $10^{-6}$ with increasing $K$) due to the high accuracy ($10-12$ digits) of the input data.  
  The second type of errors is due to the finite-size effects, and it comes 
  from neglecting the higher order correction terms in (\ref{eq:h2Conv}) and (\ref{eq:gConv}).
  They are in all cases one or more orders of magnitude larger than the numerical errors.
  In the tables, the extrapolated values are presented up to the first digit that differs from the exact value. 
  \begin{table}   
  \caption{Numerical ($h_{2, \text{extr} }$), and analytical ( $h_{2, \text{edge} }$) values for the Yang-Lee edge, followed by the 
           numerical convergence exponent $y_{2}$ and its exact value $y_h$ from Table \ref{tb:yhygq2}.
           The error bars are estimated to be of order of the last cited digit. The numerical part of the errors is at least 
           one order of magnitude smaller than the overall errors (including also the finite-size corrections).}
  \label{tb:q2hedgeExtrImh}
  \begin{ruledtabular}
  \begin{tabular}{cclclclcccc}
  $K$ & & $\;\;\;h_{2, \text{extr} }$    &  & $\;\;\;h_{2, \text{edge} }$ & & $\;\;\;y_2$ & &$y_h$   \\  
  \hline  
  2.5 & & $\;\;\;10^{-9}$        &  &  $0$                  & & $1.00002$ & &  $1$     \\ 
  2.0 & & $\;\;\;10^{-8}$        &  &  $0$                  & & $0.7497$  & &  $3 / 4$ \\
  1.9 & & $\;\;\;0.01514$        &  &  $0.015136\ldots$     & & $0.664$   & &  $2 / 3$ \\  
  1.5 & & $\;\;\;0.1814$         &  &  $0.18117 \ldots$     & & $0.65996$ & &  $2 / 3$ \\ 
  1.1 & & $\;\;\;0.478$          &  &  $0.4756\ldots$       & & $0.67$    & &  $2 / 3$\\ 
  \end{tabular}
  \end{ruledtabular}
  \end{table} 
  Results presented in the second column clearly indicate the opening of the gap for $K < K_0 = 2$.
  Good agreement also exists between the convergence exponent $y_2$ and the exactly known anomalous dimension 
  $y_h$ of the field (fourth and fifth columns). 

  The second quantity of interest is the edge density of zeros, $g_N$, defined as 
  \be  \label{eq:gN}
  g_N = \frac{1}{N}\;\frac{d\,n}{d\,l}. 
  \ee
  The density of zeros near the Yang-Lee edge can be numerically calculated from the distance between the two zeros
  closest to the real axis, $h^{(0)}$ and $h^{(1)}$,
  \be \label{eq:dl12}
  d\,l = \left[ \Big(h_1^{(1)} -h_1^{(0)} \Big)^2 +  \Big(h_2^{(1)} -h_2^{(0)}\Big)^2 \right]^{1/2},   
  \ee 
  using a discrete version of Eq. (\ref{eq:gN}) with $d\,n = 1$.
  It was extrapolated to the limit $N \rightarrow \infty$ by a least-squares fit to the form given by Eq. 
  (\ref{eq:defyg}), i.e., 
  \be  \label{eq:gConv}
   g_{N} = g_{\text{extr}} + \frac{\text{const}}{N^{y_{g, \text{extr} } } }.
  \ee

  The results of these extrapolations are presented in Table \ref{tb:q2hedgeExtrg}.
  \begin{table}   
  \caption{Numerical ($g_{\text{extr} }$) and analytical [$g_{\text{exact} }$, from Eq. (\ref{eq:DeltaSg})] values for the edge density 
           of zeros, followed by the numerical exponent $y_{g, \text{extr} }$ and its exact value $y_g$ 
           (from Table \ref{tb:yhygq2}).
           The error bars are estimated to be of order of the last cited digit. The numerical part of the errors is at least 
           one order of magnitude smaller than the overall errors (including also the finite-size corrections).}
  \label{tb:q2hedgeExtrg}
  \begin{ruledtabular}
  \begin{tabular}{ccccccccccc}
  $K$ & & $\;\;\;\;\;\;g_{\text{extr} }$   &  &\;\;\;\;\;\;$g_{\text{exact} }$& & $y_{g, \text{extr} }$       & &  $y_g$ \\   \hline 
  2.5 & & $\;\;\;0.11306554704$    &  &$0.11306554697$        & & $0.9999$    & &  $1$           \\  
  2.0 & & $10^{-6}$                &  & 0                     & & $0.2501$    & &  $1 / 4$       \\
  1.9 & & $10^{-3}$                &  & 0                     & & $0.29\;\;\;$& &  $1 / 3$       \\
  1.5 & & $10^{-3}$                &  & 0                     & & $0.30\;\;\;$& &  $1 / 3$       \\
  1.1 & & $10^{-2}$                &  & 0                     & & $0.28\;\;\;$& &  $1 / 3$  
  \end{tabular}
  \end{ruledtabular}
  \end{table} 
  The vanishingly small numerical values of the edge density of zeros  indicate a second-order transition for
  $K \leq 2$. At lower temperatures, the transition is of first order, and the edge density of zeros has a finite value.
  The critical exponent of the Yang-Lee edge singularity is 
  in excellent agreement with the exact value for the Ising phase transition, given by the scaling at $K=2$. 
  The scaling exponent for $K \leq 2$, corresponding to the Yang-Lee edge singularity, indicates a different critical 
  regime, but is obtained with less precision.

  \section{$q = 3$ model} \label{sec-q3}

  For the three-state Potts model in the symmetry breaking field, the reduced energy may be written as 
  \be \label{eq:Energijaq3}
  -\,\frac{E}{k_B\,T}  =  \frac{K}{N}\;\left[ \binom{N_1}{2} +  \binom{N_2}{2}  +  \binom{N_3}{2}  \right] + h\; N_1,
  \ee 
  \be
  N_1 + N_2 + N_3 = N, 
  \ee
  where $N_1, N_2, N_3$ denote numbers of particles in the three respective states.
  The external field is conjugated here to the Potts state $1$ and (dis)favors it, depending on the sign of $h$.

\begin{figure}
  \includegraphics[width=0.9\linewidth]{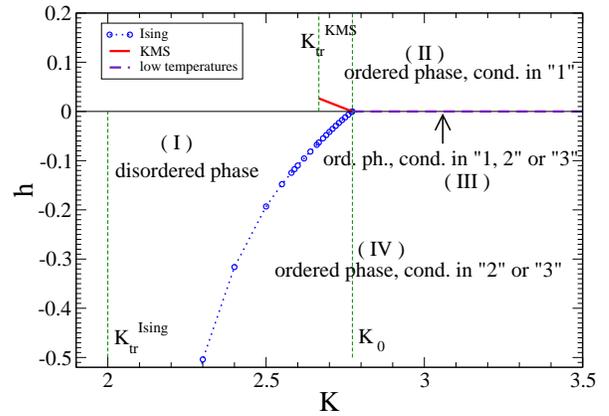}     
  \caption{(Color online) Phase diagram: Ising transition ($h < 0$), open circles; KMS transition ($h > 0$), thick full line.
          At low temperatures, $K > K_0 (3)$, Ising and KMS lines are merged and a transition exists only for 
         $h = 0$, broken line.
         \label{fg:FazniDijagKh}}
  \end{figure}

 Compared to the Ising case, this model has a much richer phase diagram (Fig. \ref{fg:FazniDijagKh}), the details of  which will be 
 discussed later in this section. 
  In addition to the number of exact results for $h \geq 0$ and arbitrary value of $q$, \cite{W82}, we pay special attention to the less
  investigated $h < 0$ part of the diagram. 
 
  For $h = 0$ and $q > 2 $, the Potts model has a first-order phase transition at inverse temperature
  \be  \label{eq:K0q}
  K_0 (q) =  2\;\dfrac{q-1}{q-2}\;\ln (q-1)
  \ee
  For $q=3$, $K_0 (3) = 4 \; \ln \, 2 = 2.772588722239781\ldots$. 

  While in the presence of an external magnetic field, there is no transition for $q =2$, models with $q > 2$
  conceal a 
  more complicated critical behavior, \cite{CET05,BCC06,BGRW08,GRW08}, with two types of field-driven 
  transitions: for $h > 0$ and for $h < 0$. \\
  For $h > 0$, there is a line of first-order transitions in $(K, h)$ plane 
  \be \label{eq:hKKMS}
  h (K) = \frac{1}{2}\;\frac{q-2}{q-1} \Big( K_0 (q) - K\Big), \hspace{0.5cm}  K_{tr} < K \leq K_0 (q),
  \ee  
  which starts at $(K_0 (q), 0)$ and ends at the tricritical point $(K_{tr},h_{tr})$ with 
  \be \label{eq:TriCritKMS}
  K_{tr} (q) = 4\,\frac{q-1}{q},  \hspace{0.5cm}  h_{tr} (q) =\ln (q-1) - 2\,\frac{q-2}{q},
  \ee
  where the transition is of second order. 
  For $q=3$, $K_{tr} (3) = 8/3 = 2.\dot{6}$, $h_{tr} (3) = 0.02648051389327864\ldots$ and the transition line is given 
  by thick full line in Fig. \ref{fg:FazniDijagKh}.\\
  For negative values of $h$, the line of first-order transitions, also starts at $(K_0 (q), 0)$ and approaches
  asymptotically to the point $(K_0 (q-1), -\infty)$ (open circles in Fig. \ref{fg:FazniDijagKh}). 
  The exact functional dependence $h = h (K)$ is not known.

\begin{figure*}
  \includegraphics[scale=0.65]{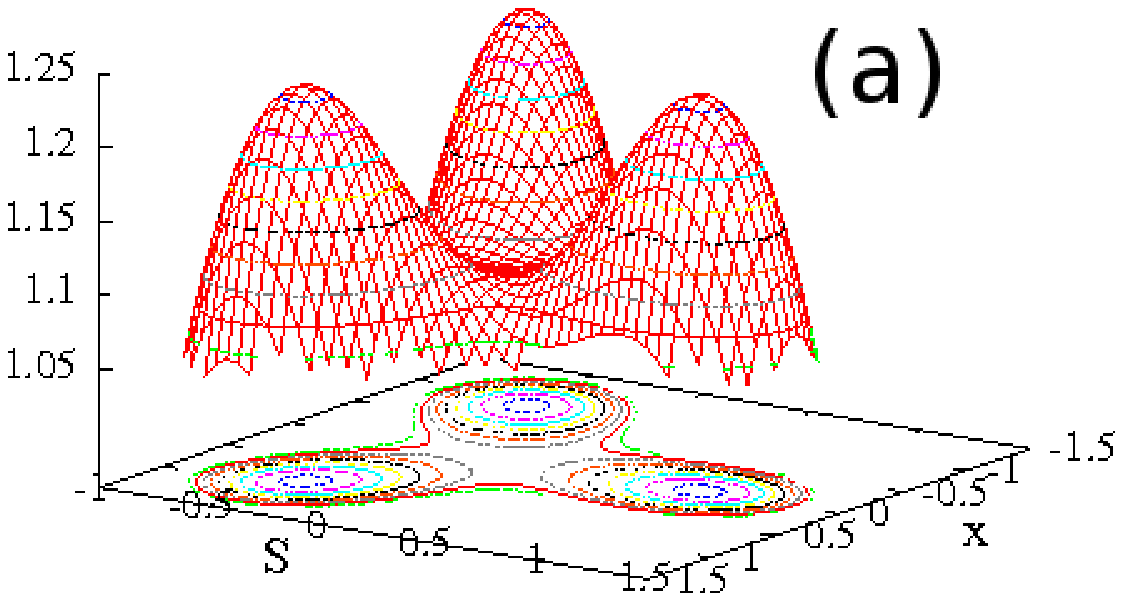}
  \includegraphics[scale=0.65]{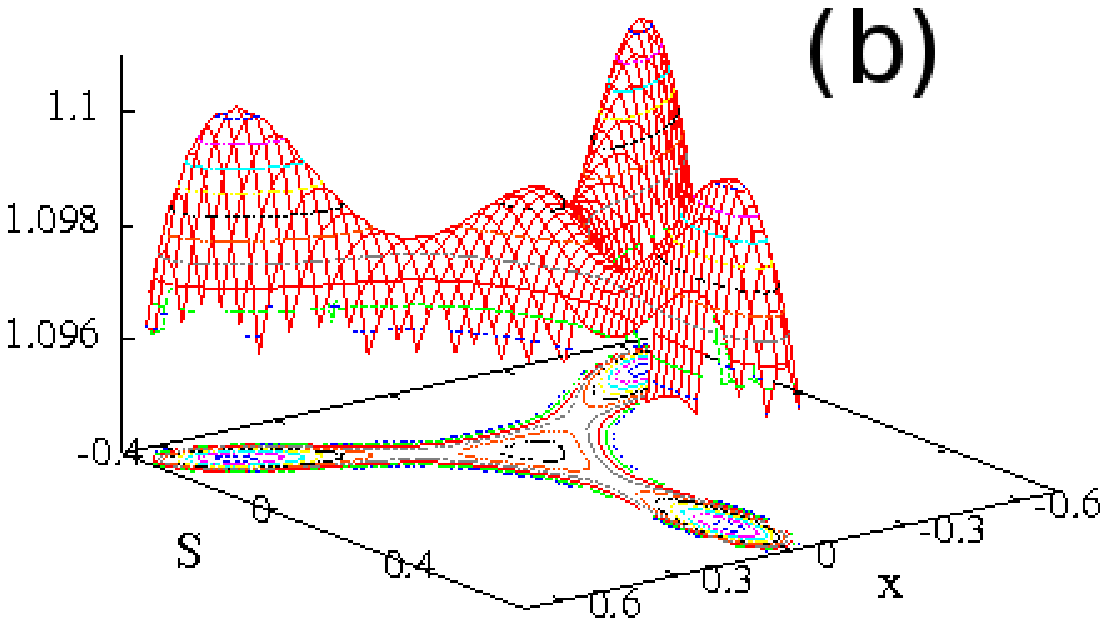}  
  \caption{(Color online) The free-energy density $- f(x, S)$ at (a) $(K, h) = (3.5, 0)$  and (b) $(K, h) = (2.79, 0)$.\label{fg:gnuplotK2t790}}
  \end{figure*}

\begin{figure} 
  \includegraphics[width=1\linewidth]{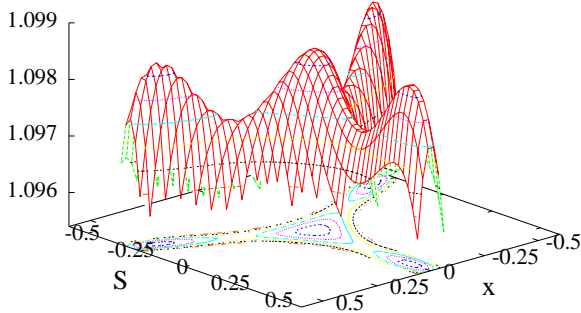}
  \caption{(Color online) The free-energy density $- f(x, S)$ at inverse temperature $K = K_0 (3) = 4\,\ln 2$ and $h = 0$. \label{fg:K2t772h0t000}}
  \end{figure}

\begin{figure*}
  \includegraphics[scale=0.65]{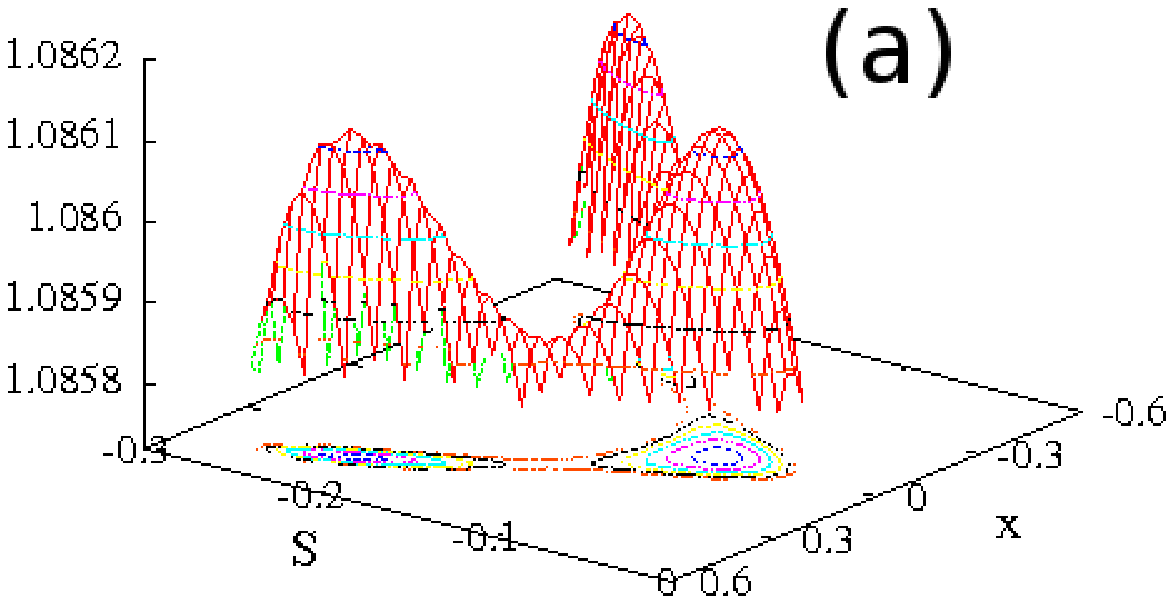}
  \includegraphics[scale=0.65]{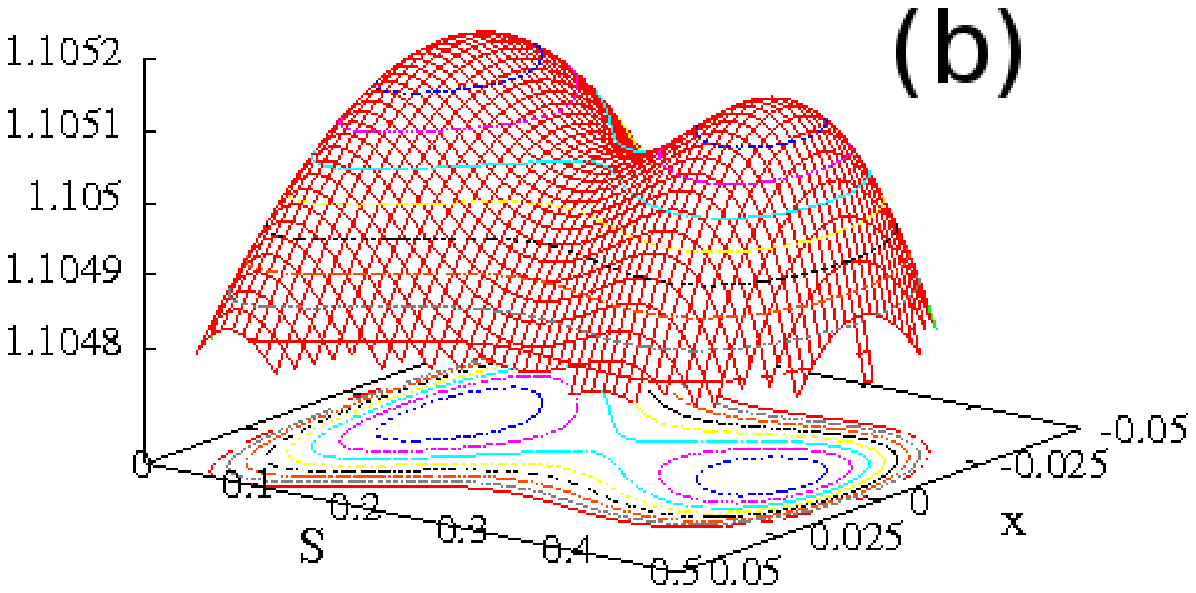}  
  \caption{(Color online) The free-energy density $- f(x, S)$ at $K = 2.70$ and two values of the field: 
          (a) $h_{Ising} = -0.04169621964961\cdots$ and (b) $h_{KMS} = 0.0181471805599452\cdots$.\label{fg:gnuplotK2t700}}
  \end{figure*}

\begin{figure*}
  \includegraphics[scale=0.65]{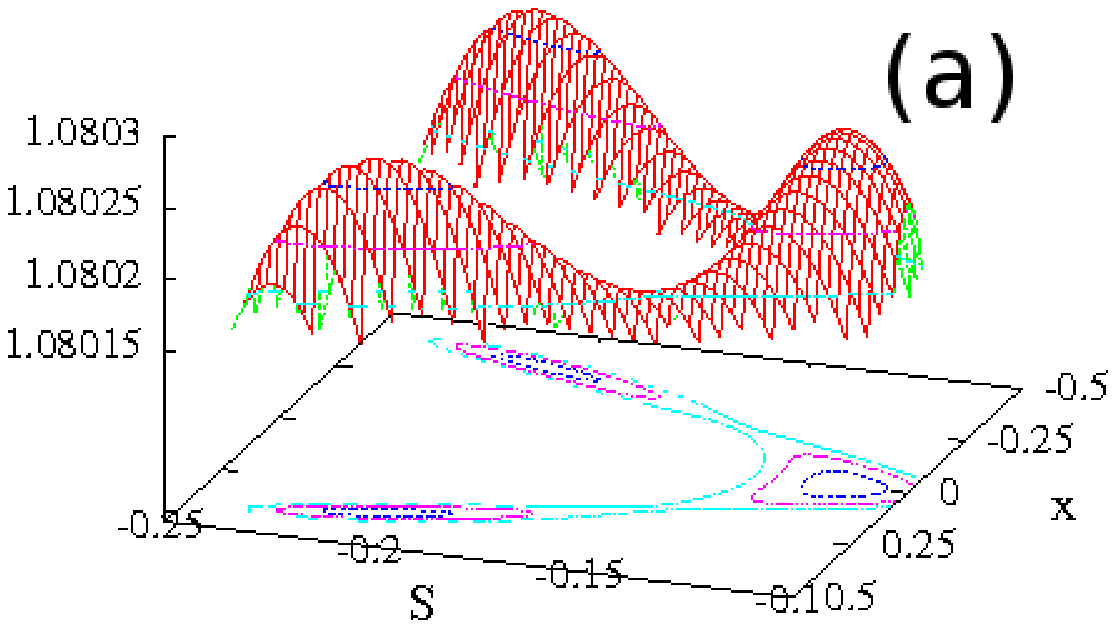}  
  \includegraphics[scale=0.65]{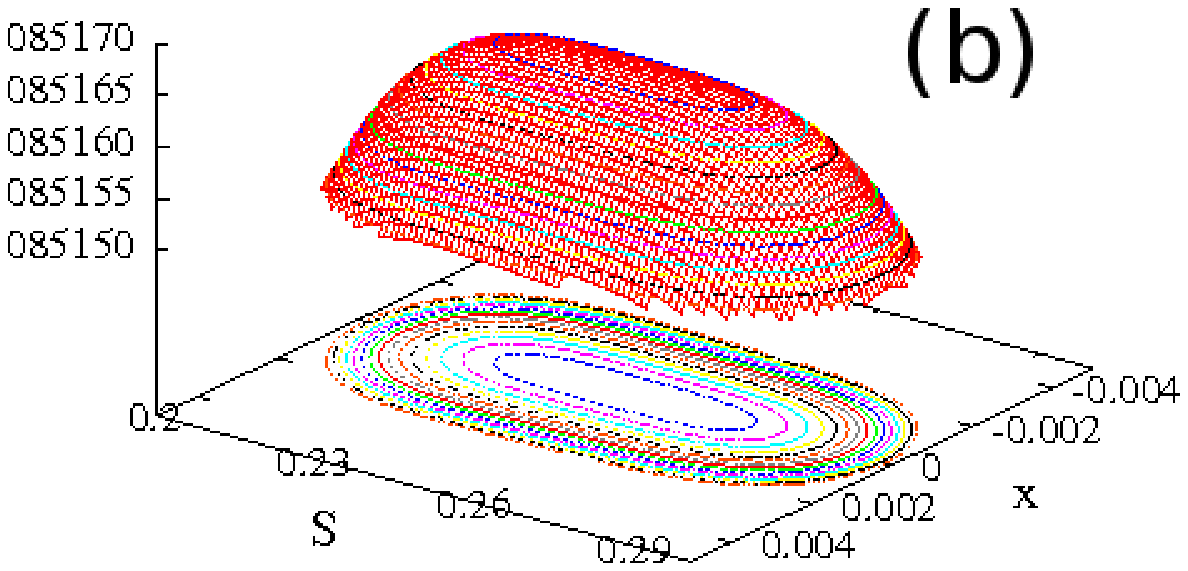}   
  \caption{(Color online) The free-energy density $- f(x, S)$ at inverse temperature $K = K_{tr}^{KMS} = 8/3 = 2.\dot{6}$ and external 
          field: (a) $\;\;h =  -0.06333826060\cdots$ and (b) $\;\;h = h_{tr}^{KMS} =  0.02648051389\cdots$.\label{fg:gnuplotK2t666}}
  \end{figure*}

   For the purpose of numerical calculations, it is useful to present the partition function in a polynomial form 
  \be \label{eq:ZNq3opcenito}
   \cZ_N = \sum_{n = 0}^N\;a_{n} (K) \;z^{\,n}, \hspace{1cm} z = e^h ,
  \ee
  with the coefficients $a_n$ given by
  \bea  \label{eq:KoefCijelePF}
  a_{n} ( K ) & = & e^{\,K (N-1) / 2} \; \frac{N\,!}{(N-n)\;! }\; e^{(K/N)[n^2-N n] }  \\
             & \cdot &\sum_{m = 0}^{n}\;\frac{ e^{(K/N)[m^2 - n \, m]}}{\;m\;!\;\;(n - m)\,! }. \nonumber 
  \eea 
  On the other hand, for the analytical approach, the partition function, written as 
  \bea \label{eq:ZNq3opcenitoSH}
   \cZ_N & = & e^{\,K (N-1) / 2} \;e^{-( N\,K / 4) [(4/3)(h/K-1/2)^2+1]}\nonumber  \\
   & \cdot & \sum_{N_1 = 0}^N\;\binom{N}{N_1}\;\sum_{N_2 = 0}^{N-N_1}\; \binom{N-N_1}{N_2}\\
   & \cdot &  e^{\,A_1^{\,2}  }\;\;e^{\, A_2^{\,2} }, \nonumber 
  \eea
   with
  \bea
  A_1 & = & \sqrt{N\,K}\;\left( \frac{1}{2}\, \frac{N_1}{N} + \frac{N_2}{N} - \frac{1}{2}\right), \nonumber \\
  A_2 & = & \sqrt{\frac{3\,N\,K}{4} }\;\left[\frac{N_1}{N} + \frac{2}{3}\;\left(\frac{h}{K}-\frac{1}{2} \right)\right],  \nonumber
  \eea
  allows a twofold application of the Hubbard-Stratonovich transformation (\ref{eq:Gauss})
  which reduces it to a double integral
  \be\label{eq:PFq3}
  \cZ_N  \; \sim\;\int_{-\,\infty}^{+\,\infty}\;\int_{-\,\infty}^{+\,\infty}
  \; d\,x\;d\,S\;\;e^{\,-N\,f(x, S)}    , 
  \ee
  with the exact free-energy density $f(x, S)$
  \be  \label{eq:GaussGustSlobEnerg}
  - f(x, S) = - K \left(\frac{x}{2} \right)^2 -\frac{K}{3} S (S + 1) + \ln \Big[ 2 \cosh \frac{x K}{2} + e^{\,K S + h}\Big],
  \ee
  where $S = ( 3\,N_1/N - 1)/2$ is the order parameter for $h \geq 0$, and $x = (N_2 - N_3)/N$ is the order 
  parameter for  $h < 0$. 
  The integral (\ref{eq:PFq3}) is then solved by a saddle-point approach.
  The locations of minima are the solutions of a two-by-two system of equations for $x$ and $S$
   \be\label{eq:ParcfPox}
  x =  2\;\tanh (x K / 2)\;\;\; \dfrac{1}{\dfrac{e^{\,K S + h} \;}{\cosh ( x K / 2) }\;\;\;+ \;\;\;2  }, 
  \ee
  \vspace{0.3cm}
  \be  \label{eq:ParcfPoS}
  S = 
  \dfrac{\dfrac{e^{\,K S + h}\;}{\cosh ( x K / 2) } \;\;\; - \;\;\;1}{\dfrac{e^{\,K S + h} \;}{\cosh ( x K / 2) }\;\;\;+ 
   \;\;\;2} .
  \ee
  The MF approaches to the general $q$-state Potts model, such as the solution by Kihara \etl  \cite{KMS54} usually 
  neglect the fluctuations among states orthogonal to the ordered one, which corresponds to taking $x = 0$. 
  We examine here the free energy in the entire $(x,S)$ plane.   
  In Figs. \ref{fg:gnuplotK2t790} - \ref{fg:gnuplotK2t666} we illustrate the shapes of $f$ for several characteristic
  values of temperature and the field, corresponding to different phases.
  For clarity in the figures, the maxima of $-f$ are displayed instead of the minima of $f$. \\
  At low temperatures, the three maxima of the same height are obtained only for $h = 0$ [Fig. \ref{fg:gnuplotK2t790}(a)].
  They correspond to a triply degenerate ordered phase of the system for $h = 0$.  
  By increasing temperature the maximum at the origin (Fig. \ref{fg:gnuplotK2t790} b), which corresponds to the disordered
  state, appears and starts to rise.  
  For $K_0 (3) = 4\,\ln 2 = 2.772\ldots$, all the four maxima reach the same height (Fig. \ref{fg:K2t772h0t000}),
  indicating the coexistence between disordered and ordered phases at the first-order transition point
   $(K, h) = (K_0 (3), 0)$.
  At still higher temperatures, $K \, < \,K_0 (3)$, the maxima behave differently for positive and negative values of $h$ 
  (Fig. \ref{fg:gnuplotK2t700}). 
  For positive values of the field, two maxima appear [Fig. \ref{fg:gnuplotK2t700}(b)], which represent the transition that
  we denote here as the KMS\footnote{After the approximation of Kihara \etl, as explained in Sec. \ref{sec-KMStransitions}.} transition,
  while for negative values of the field, there are three maxima [Fig. \ref{fg:gnuplotK2t700}(a)] representing the 
  transition denoted here as Ising\footnote{See Sec. \ref{sec-IsingTransitions}.} transition. \\
  At the KMS transition, the distance between the two maxima diminishes as $K$ goes to its tricritical value, 
  until they eventually merge at $(K_{tr}^{KMS}, h_{tr}^{KMS})$ given by (\ref{eq:TriCritKMS}) 
  [Fig. \ref{fg:gnuplotK2t666}(b)].\\
  The transition denoted as Ising, exists for all $K_0 (3) > K > K_{tr}^{Ising} = K_0 (2) = 2$. 
  At $K_{tr}^{Ising}$ and $h_{tr}^{Ising} = -\infty$,  all three maxima [similar to those from the 
  Fig. \ref{fg:gnuplotK2t666}(a)], 
  merge and the system undergoes a second-order phase transition of an Ising universality class.\\
 For a summary of the critical behavior of the present model let us turn back to the phase diagram in Fig. \ref{fg:FazniDijagKh}. 
 There are four distinct phases: \\
  (I) Disordered phase at high temperatures, \\
  (II) For $h>0$, there is an ordered phase in the Potts state $1$. Ground state of this phase is non degenerate. 
       Transition between this phase and disordered phase is of first order (the thick full line in Fig. \ref{fg:FazniDijagKh}), 
       except at the end point (tricritical point), where the transition is of second order, and where it is possible 
       to go continuously from the phase I to the phase II. \\
  (III) For $h=0$, there is a line of first-order transitions between the phase II and the phase IV (broken line in Fig.
        \ref{fg:FazniDijagKh}). Its ground state is triply degenerate. \\
  (IV) For $h<0$, the number of particles in state $1$, coupled to the field, is suppressed favoring the  particles
       in the Potts states $2$ or $3$. Ground state of this phase is twice degenerate. 
       The first-order transition line separating the phase IV and the disordered phase is marked by the open circles in
       Fig. \ref{fg:FazniDijagKh}. This line has no endpoint and goes down to $h = -\infty$. The continuous transition
       between phases IV and the disordered phase is not possible, except at infinity.

 Let us discuss first the scaling properties at different transitions.

  \subsection{$h > 0$, KMS transitions} \label{sec-KMStransitions}

  A set of solutions satisfying the extrema conditions (\ref{eq:ParcfPox}) and (\ref{eq:ParcfPoS}) corresponds to $x = 0$,
  while $S$ is the solution of the equation
  \be        \label{eq:KMSextrS}
  S  =  \frac{e^{ K S + h} - 1}{ e^{ K S + h} + 2}  .
  \ee 
  The same equation is found in the approximation used by Kihara \etl \cite{KMS54,BCC06} to the second order phase
  transition in the MF Potts model,  which neglects the fluctuations within the remaining $(q-1)$ states.
  For $h$ and $K$, related through the Eq. (\ref{eq:hKKMS}) (the thick full line in Fig. \ref{fg:FazniDijagKh})
  \be \label{eq:KMShK}
  h = \ln \,2 -  \frac{K}{4},   
  \ee
  the positions of the extrema $S_{\pm}$ [Fig. \ref{fg:gnuplotK2t700}(b)] are of the form
  \be
  S_{\pm} =  \frac{1}{4}\; \pm \; \Delta\; S,  
  \ee
  where $\Delta\; S$ is a solution of the equation
  \be  \label{eq:DeltaS}
  K\; \Delta\;S  =   \ln \; \frac{1+ \frac{4}{3}\;\Delta\;S}{1- \frac{4}{3}\;\Delta\;S} .
  \ee
  As the temperature rises, $K < K_0 (3)$, the distance between the two maxima $S_{\pm}$ decreases and eventually
  vanishes at the tricritical point (\ref{eq:TriCritKMS})
  \be
  K_{tr}^{KMS} = \frac{8}{3},  \hspace{2cm} h_{tr}^{KMS} = \ln \, 2 - \frac{2}{3}
  \ee
  (endpoint of the thick full line in Fig. \ref{fg:FazniDijagKh}).

  In the first-order transition regime and for large values of $N$, the dominant contribution to the partition function (\ref{eq:PFq3}) 
  comes from the two minima of equal depth
  \bea      
  Z_N & \sim & e^{\;\;- N\; f (S_+)  } +  e^{\;\;- N\; f (S_-)  } \nonumber \\
      & \sim & e^{\;\;- N\; f (S_+)  } \left\{ 1 + e^{\;\;- N\; [  f (S_-) - f (S_+) ]   }\right\}  .
  \eea  
  Within this approximation, the calculation of zeros in the above expression, reduces to the following set of equations for the free-energy densities at $S_{\pm}$ 
  \bea 
  Re \Big[f (S_-) - f (S_+)  \Big] & = & 0,  \nonumber  \\  \\
  N \,\cdot \, Im  \Big[f (S_-) - f (S_+)  \Big] & = & (2 n + 1) \, \pi, \hspace{0.5cm} n = 0, 1, \cdots  \nonumber
  \eea
  To the leading order in $1/N$, the solutions of the above equations are of the form
  \bea  \label{eq:nuleanalit} 
  h_1^{(n)} & = & \ln 2 -  \frac{K}{4} - \frac{1}{N}\;\frac{3\,K}{8} + \cO \left(\frac{1}{N^2} \right), 
     \nonumber \\
  & & \\
  h_2^{(n)} & = & \frac{1}{\Delta\;S}\;\frac{3}{4}\; \frac{2 n + 1}{N}\; \pi  + \cO \left(\frac{1}{N^2} \right). \nonumber
  \eea
  Remark that, to  the leading order in $1/N$, the real part of zeros does not depend on $n$. 
  Numerical calculations (Figs. \ref{fg:YLMF-q3N3000K2t62t7}  and  \ref{fg:YLMF-q3N10000K03t5}) suggest that this remains
  true to all orders in  $1/N$, i.e. that, at KMS transition, all the zeros lie on a straight line parallel to 
  the imaginary axis.

  Calculation of the density of zeros at the Yang-Lee edge is performed in the same way as for the Ising $(q = 2)$ case. 
  Relation (\ref{eq:gN}) defines the density of zeros, while its value at the edge is given by the distance between the two zeros closest to the real axis.
  By inserting expansions (\ref{eq:nuleanalit}) into (\ref{eq:dl12}), one gets 
  \be  \label{eq:gexact}
   g_N = \frac{2\; \Delta\;S}{3\,\pi}  + \cO \left(\frac{1}{N} \right).
  \ee
  On the entire first-order transition line, $\Delta\;S > 0$ making $g$ finite. By approaching the tricritical point, 
  $\Delta\;S \to 0$, and vanishing of $g$ indicates the second-order transition at the tricritical point.

  \subsection{$h < 0$, Ising transitions} \label{sec-IsingTransitions}
 
  Negative values of the field bring the particles in the Potts state $1$ (as defined in Eq. (\ref{eq:Energijaq3}) )
  to  an energetically higher level than the particles in the states $2$ and $3$. 
  Consequently,  the system prefers to have most of the particles in states $2$ and $3$. In the limit 
  $h \, \to \, - \infty$, transitions of particles into state $1$ are completely forbidden and the model reduces to
  the pure two-state (Ising) model without an external field. Thus, the negative field 
  acts as a chemical potential: it regulates the number of particles in states $2$ and $3$.

  In the limit $h \, \to \, - \infty$  the extrema conditions (\ref{eq:ParcfPox}, \ref{eq:ParcfPoS}) and the free-energy
  density (\ref{eq:GaussGustSlobEnerg}), reduce to
  \be \label{eq:IsingSolution}
  x  =  \frac{e^{x\, K}-1}{e^{x \,K} +1  }, \hspace{2cm}
  S  = -\,\frac{1}{2} .
  \ee
  \be
  - f(x, - 1 / 2) =  - K \left(\frac{x}{2} \right)^2  + \ln \Big[ 2 \cosh \frac{x K}{2}\Big]  + const.
  \ee
  Equation (\ref{eq:IsingSolution}) has $x \, \to \, 0$ solution at $K = K_{tr}^{Ising} = K_0 (2) = 2$, so that the 
  position of Ising tricritical point is
  \be
  \Big( K_{tr}^{Ising}, h_{tr}^{Ising} \Big) = (2, -\,\infty) .
  \ee
  In the range of inverse temperatures $K_0 (2) < K < K_0 (3)$ (open circles in Fig. \ref{fg:FazniDijagKh}), 
  the free-energy density has a shape similar to the one shown in Fig. \ref{fg:gnuplotK2t666}(a), and the transition is of
  first order. 
  By approaching $K \, \to \, K_{tr}^{Ising} = 2$, the first order character of the transition becomes weaker and, 
  at $K= 2$, all three extrema of the free energy merge, and the transition changes its character into a second-order
  one. 

  By studying the behavior of the free-energy density around the two edges of the Ising line of transitions, 
  it is possible to calculate the shape of $h = h (K)$ in the two limits. 
  Close to the point $(K, h) = (K_0 (3), 0) $
  \be
  h (K) \simeq \ln\;  \Big(\; e^{\,\frac{K}{4}} -1 \;\Big).
  \ee
  Close to the point $(K, h) = (K_0 (2), -\,\infty) $.
  \be
  h (K) \simeq \ln\;  \Big(\;  K - 2 \;\Big) \,+ \,\frac{K}{2}.
  \ee

  In reference \cite{BCC06}, authors discuss the cases with $h < 0$ and $q \geq 4$ proving the existence of the
  transition in the range
  \be
  K \;\in \; \Big(K_0 (q-1), K_0 (q) \Big) \;\; , \;\;  h \;\in \; \Big(-\,\infty, 0 \Big) .
  \ee
  Our results show explicitly how the above statement is extended to the case $q = 3$.

  \subsubsection{Numerical results} \label{sec-q3Numeric}

\begin{figure*}
  \includegraphics[scale=0.35]{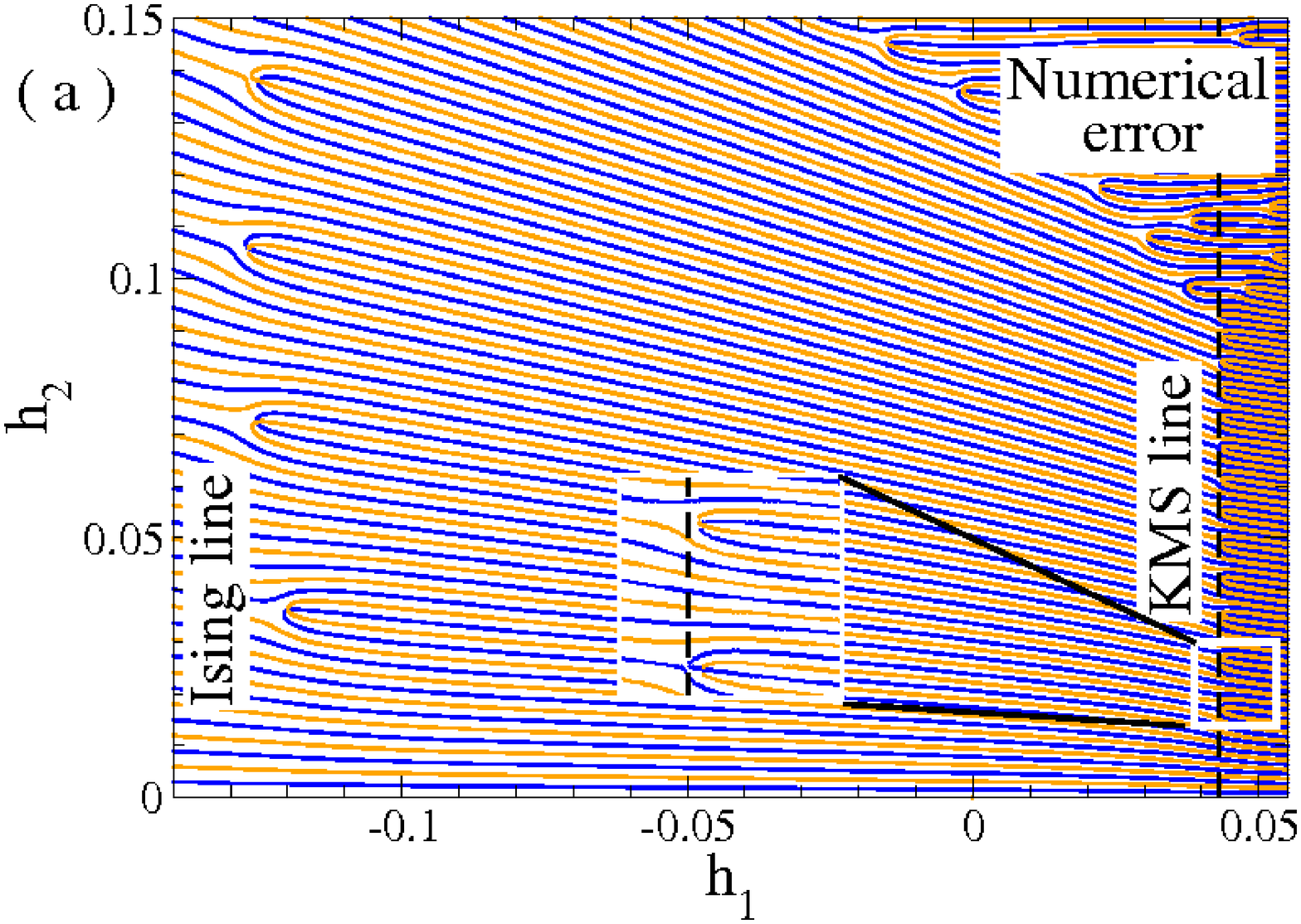}  
  \includegraphics[scale=0.35]{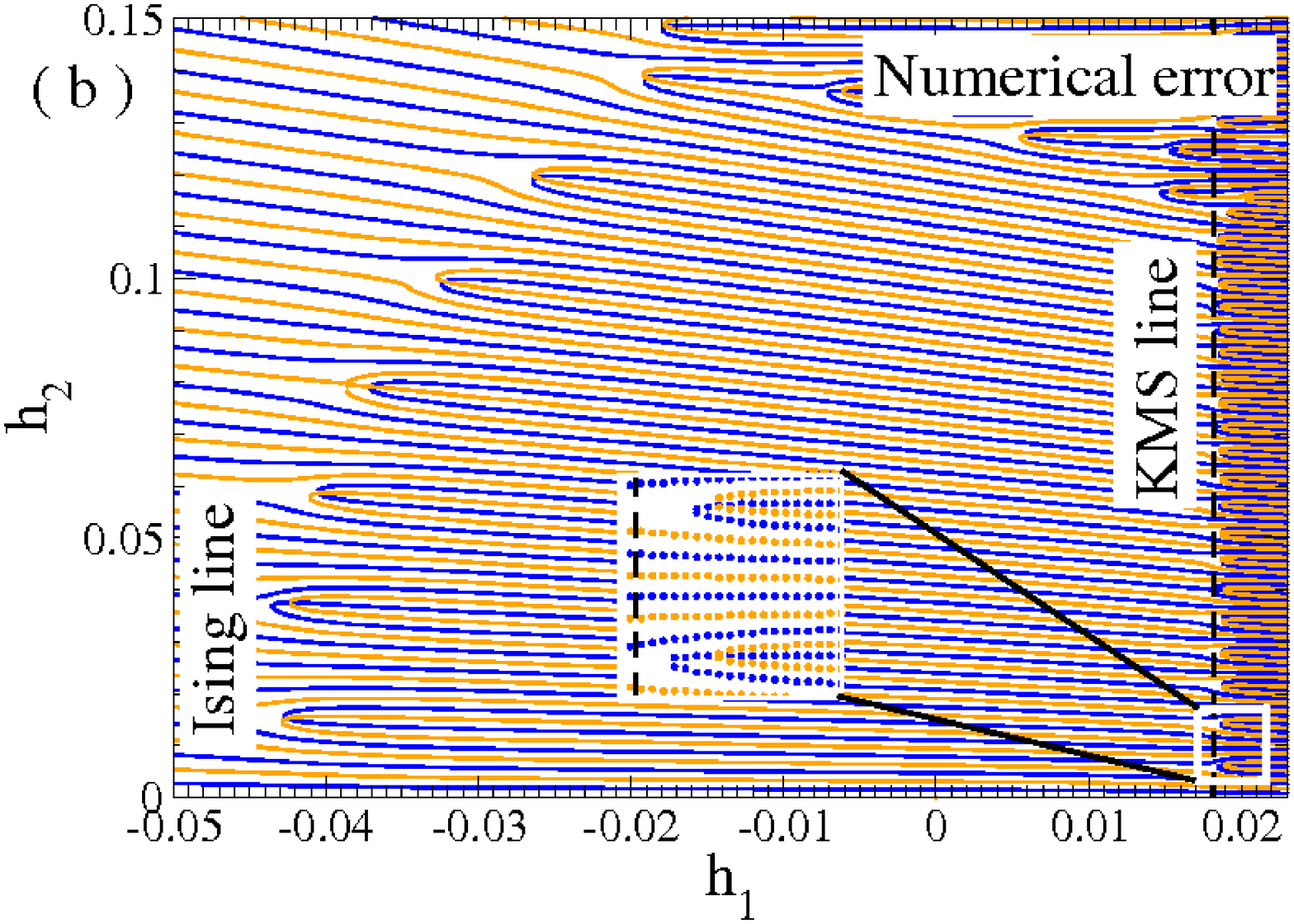} 
  \caption{(Color online) $q = 3$ and $ N = 3\,000$. The positions of the part of the Yang-Lee zeros 
          closest to the real field axis are given as the intersections of the blue (dark gray) and orange (light gray) lines:
          (a) $K = 2.6$ and (b) $K = 2.7$. The broken black line is the KMS value of the real field predicted by relation
          (\ref {eq:KMShK}).\label{fg:YLMF-q3N3000K2t62t7}}
  \end{figure*}

\begin{figure*}
  \includegraphics[scale=0.34]{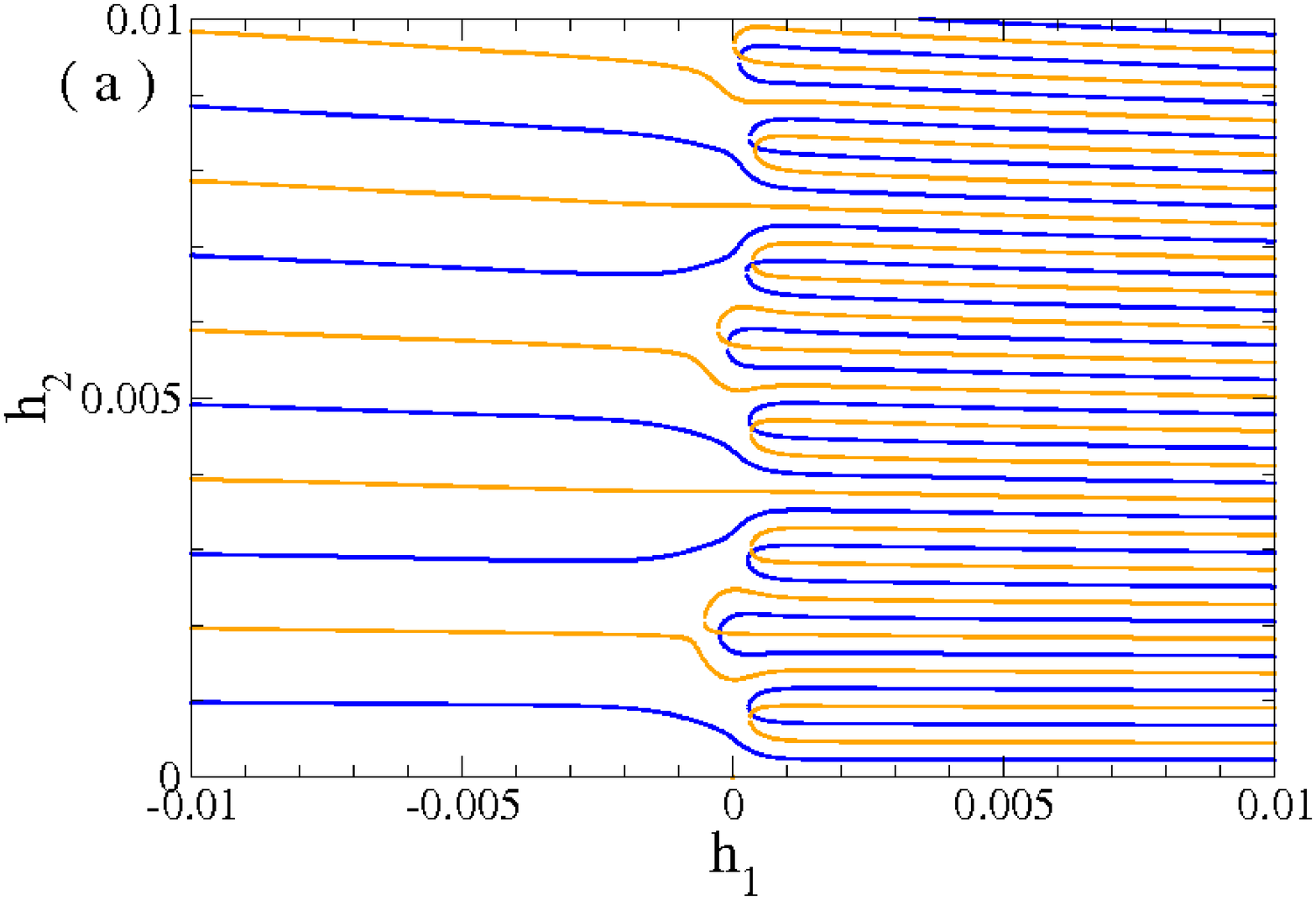} 
  \includegraphics[scale=0.34]{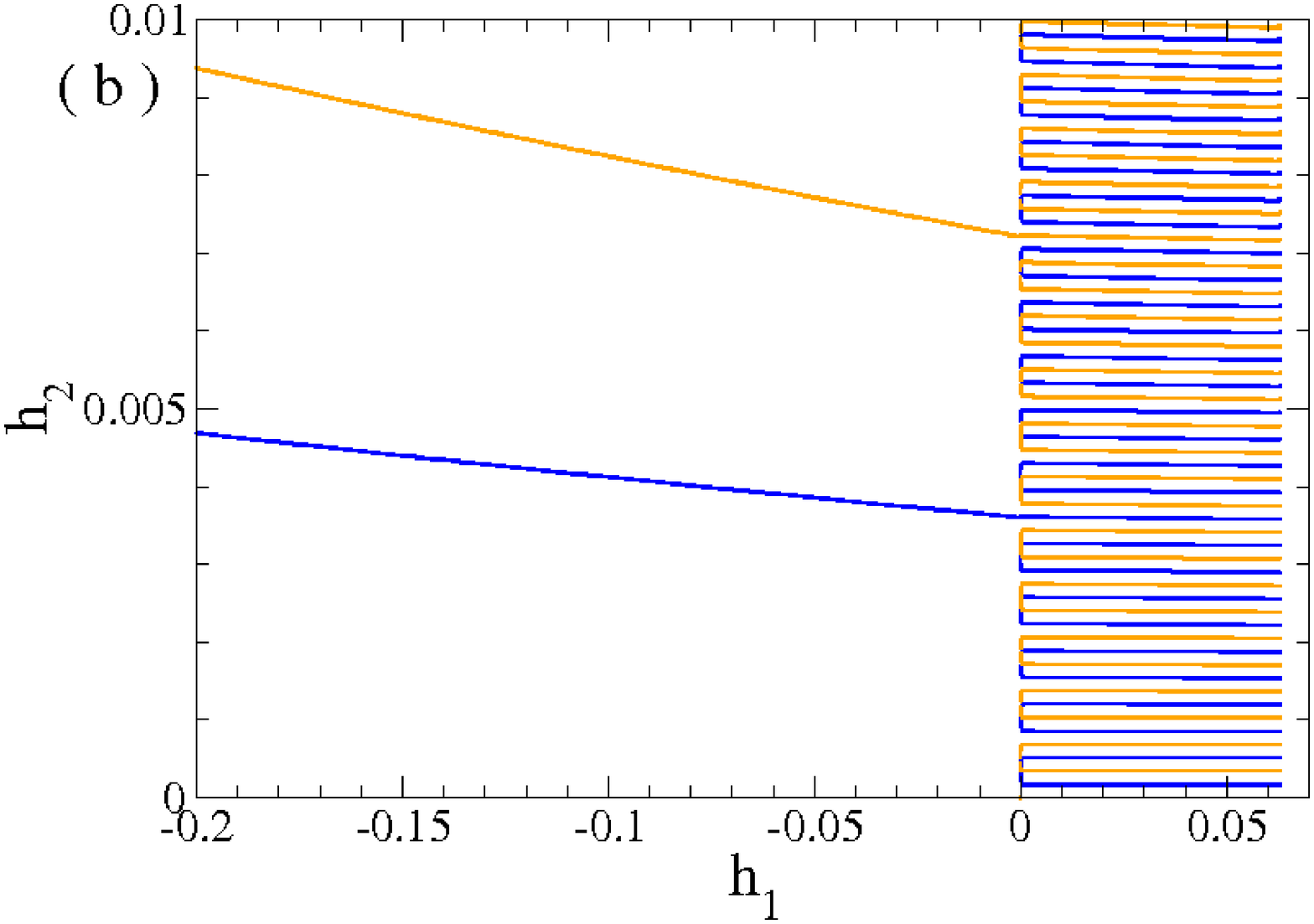} 
  \caption{(Color online) $q = 3$ and  $N = 10\,000$. The positions of the part of Yang-Lee zeros 
          closest to the real field axis are given as the intersections of the blue (dark gray) and orange (light gray) lines:
         (a) $K = K_0 (3) = 2.772\ldots$ and  (b) $K = 3.5$.\label{fg:YLMF-q3N10000K03t5}}
  \end{figure*}

  The numerical calculation of Yang-Lee zeros starts from Eq. (\ref{eq:ZNq3opcenito})
  \be 
   \cZ_N = \sum_{n = 0}^N\;a_{n} (K) \;z^{\,n} = Re\;\cZ_N + \imath\; Im \; \cZ_N = 0.
  \ee
  Numerical solutions of $Re\;\cZ_N = 0$ [blue (dark gray) lines] and $Im\;\cZ_N = 0$ [orange (light gray) lines]
  with their intersections giving
  the positions of Yang-Lee zeros are illustrated in Figs. \ref{fg:YLMF-q3N3000K2t62t7} and \ref{fg:YLMF-q3N10000K03t5} 
  for two typical values of $K$.
  Two lines of zeros, similar to those in Fig. \ref{fg:YLMF-q3N3000K2t62t7}, were also observed 
  in a numerical study \cite{BBCKK00}, of a three-dimensional $q = 25$ Potts model with nearest-neighbour interactions
  on small lattices. \\

  Following  the same procedure as the one presented in Sec. \ref{sec-q2}, we analyzed the positions of the two zeros closest to 
  the real axis for a sequence of large system sizes ranging from $N=10^{\,3}$ up to $10^{\,6}$ sites.
 The  position of the edge of the line of zeros and the edge density of  zeros were extrapolated to the limit $N\,\to\,\infty$ by using
  the simple least-squares fit to the form \be
   h_{1, 2, N }  =  h_{1, 2, \text{extr} } + \frac{\text{const} }{N^{y_{1, 2}}}  \;\; , \;\; 
   g_{N} =  g_{\text{extr} } + \frac{\text{const} }{N^{y_g}}  .
  \ee
  The results of these extrapolations in the three different regimes, corresponding to 
  $(K < K_0 (3), h > 0), (K \geq K_0 (3), h = 0)$, and $(K < K_0 (3), h < 0)$, 
  are presented in Tables  \ref{tb:q3RehImhgExtrKMS},  \ref{tb:q3RehImhgExtrCijelaNiskeTemp}, and \ref{tb:q3RehImhgExtrCijela1Ising}.
  The $K < K_0 (3)$ regime contains two lines of transitions and in addition the Yang-Lee edge singularity
   for sufficiently high temperatures so that the gap around the real axis is open\footnote{In present work, we analyzed it only around
   $h > 0$ tricritical point.}.
  At lower temperatures, $K \geq K_0 (3)$, there is only one, $h = 0$, line of zeros.

  \begin{table*}         
  \caption{The results of extrapolation for  $h > 0$: real and imaginary parts of the closest zeros, $h_{1,2}$, and the edge density of 
            zeros $g$, followed by the corresponding convergence exponents.
            The error bars are estimated to be of order of the last cited digit. The numerical part of the errors is at least 
           one order of magnitude smaller than the overall errors (including also the finite-size corrections).
          \label{tb:q3RehImhgExtrKMS}  }
  \begin{ruledtabular}
  \begin{tabular}{cclclclcccc}
  $K$             & &$\;\;\;\;\;\;h_{1, \text{extr} }$&  & $\;\;\;\;\;\;h_{1, \text{exact} }$& & $y_{1, \text{extr} }$ & & $y_{1, \text{guess} }$ \\
                  \hline   
  $K_0 (3) =  2.772\cdots$
                  & & $\;\;\;10^{-12}$      &  &$0$                    & &$1.000003$& &$1$    \\
  $2.7$           & & $\;\;\;0.018147180556$&  &$0.018147180559\cdots$ & &$1.000003$& &$1$    \\
  $K_{tr}^{KMS} = 2.\dot{6}$
                  & & $\;\;\;0.026480513887$&  &$0.026480513893\cdots$ & &$1.000004$& &$1$   \\
     $2.6$        & & $\;\;\;0.04314719$    &  &$0.04314718\cdots$     & &$0.99993$ & &$1$  \\
     $2.5$        & & $\;\;\;0.0681470$     &  &$0.0681472\cdots$      & &$1.0005$  & &$1$  \\
     $2.4$        & & $\;\;\;0.0931470$     &  &$0.0931472\cdots$      & &$1.0006$  & &$1$  \\
     $2.3$        & & $\;\;\;0.1181471$     &  &$0.1181472\cdots$      & &$1.0005$  & &$1$  \\
                  & &                   &  &                       & &                    & &                  \\
  $K$             & & $\;\;\;\;h_{2, \text{extr} }$&  & $\;\;\;\;\;\;h_{2, \text{exact} }$ & & $y_{2, \text{extr} }$ & & $y_{2, \text{guess} }$   \\
                   \hline   
  $K_0 (3) =  2.772\cdots$
                  & & $\;\;\;10^{-8}$        &  &$0$             & &  $1.001$           & &     $1$         \\
  $2.7$           & & $\;\;\;10^{\,-7}$      &  &$0$             & &  $1.01$            & &     $1$      \\ 
  $K_{tr}^{KMS} =  2.\dot{6}$
                  & & $\;\;\;10^{-8}$        &  &$0$             & &  $0.749$           & &     $3 / 4$      \\
     $2.6$        & & $\;\;\;0.00533$        &  &$ -$            & &  $0.668   $        & &     $2 / 3$      \\
     $2.5$        & & $\;\;\;0.02126\;\;$    &  &$ -$            & &  $0.663  $         & &     $2 / 3$      \\
     $2.4$        & & $\;\;\;0.04347$        &  &$ -$            & &  $0.658    $       & &     $2 / 3$      \\
     $2.3$        & & $\;\;\;0.07091$        &  &$ -$            & &  $0.655    $       & &     $2 / 3$      \\
                  & &                        &  &                & &                    & &                  \\
  $K$             & & $\;\;\;\;\;\;g_{\text{extr} }$ &  & $g_{\text{exact} }$             &  & $\;\;y_{g, \text{extr} }$   & &  $y_{g, \text{guess} }$     \\
    \hline   %
  $K_0 (3) =  2.772\cdots$
                  & & $\;\;\;0.05305168$     &  & $0.05305164\cdots$      &  &$0.997$               & &        $1$ \\
  $2.7$           & & $\;\;\;0.03048$        &  & $0.03047\cdots$         &  &$0.94$                & &        $1$ \\
  $K_{tr}^{KMS} =  2.\dot{6}$
                  & & $\;\;\;10^{-6}$      &  &$0$                      &  &$0.2503$              & &      $1/4$   \\
    $2.6$         & & $\;\;\;10^{-3}$      &  &$0$                      &  &$0.28$                & &      $1/3$     \\
    $2.5$         & & $\;\;\;10^{-3}$      &  &$0$                      &  &$0.29$                & &      $1/3$      \\
    $2.4$         & & $\;\;\;10^{-3}$      &  &$0$                      &  &$0.335$               & &      $1/3$     \\
    $2.3$         & & $\;\;\;10^{-3}$      &  &$0$                      &  &$0.29$                & &      $1/3$      \\
  \end{tabular}
  \end{ruledtabular}
  \end{table*}     
  Table \ref{tb:q3RehImhgExtrKMS} contains data describing the $h > 0$ or KMS line of transitions (thick full line in Fig.
  \ref{fg:FazniDijagKh}). \\
  At $K \geq K_{tr}^{KMS}$, the imaginary part of the field, is equal to zero, which means that the gap is closed at
  low temperatures.
  At higher temperatures, $K < K_{tr}^{KMS}$, the zeros accumulate around the point $(h_1, h_2 > 0)$ which means that 
  the gap is
  open with the Yang-Lee edge singularity at its edge. Real part of the edge still satisfies Eq. (\ref{eq:KMShK}). \\
  Density of zeros at $K > K_{tr}^{KMS}$ has a finite value, and the transition is of first order.  
  This finite value can be compared to the analytical expression (\ref{eq:gexact}) with high accuracy. 
  At higher temperatures, $K \leq K_{tr}^{KMS}$, 
  the edge density of zeros vanishes and the transition is of second order.\\
  The set of convergence exponents $y_{2, extr}$ and $y_{g, extr}$ has the same values as the corresponding exponents of
  the $q = 2$ model (Table \ref{tb:yhygq2}), showing that it belongs to the same universality class. 
  Since within the approximation by  Kihara \etl the handling of fluctuations in the model with $q > 2$ is essentially reduced to the  
  calculation of fluctuations in the two-state model, this result is not surprising. \\
  One may clearly distinguish the tricritical exponents for $K=K_{tr}$ from the Yang-Lee edge exponents obtained for higher
  temperatures ($K<K_{tr}$).
  The convergence exponent $y_{1, extr}$ is very close to $1$, which is anticipated by Eq. (\ref{eq:nuleanalit})  
  for the first order transitions. At $K \leq K_{tr}^{KMS}$, the transition is of second order, and we have no analytical
  expression for $y_1$, but it seems that  the  value $1$ remains for all temperatures. 

  At low temperatures, $K \geq K_0 (3)$ (see Table \ref{tb:q3RehImhgExtrCijelaNiskeTemp}), zeros of the partition function are the 
  result of the competition
  between the three (at $K > K_0 (3)$, Fig. \ref{fg:gnuplotK2t790}) or the four (at $K = K_0 (3)$, Fig. \ref{fg:K2t772h0t000})
  extrema of $f$. 
  The transition is of a strong first-order type and loci of zeros form a straight line $h_1 = 0$ (Figs. \ref{fg:YLMF-q3N10000K03t5}
  a and \ref{fg:YLMF-q3N10000K03t5} b).
  The convergence exponents are equal to $1$, except for the exponent of the edge density of zeros at $K_0 (3)$.
  \begin{table*}    
  \caption{The results of extrapolation for  $K \geq K_0 (3)$, $h=0$: the real and imaginary part of
  the closest zeros, $h_{1,2}$ and the edge density of zeros $g$, followed by the corresponding convergence exponents.
  The error bars are estimated to be of order of the last cited digit. The numerical part of errors is by at least 
           one order of magnitude smaller than the overall errors (including also the finite-size corrections).
           \label{tb:q3RehImhgExtrCijelaNiskeTemp}}
  \begin{ruledtabular}
  \begin{tabular}{cccccclcccc}
   $K$              & & $h_{1, \text{extr} }$           &  & $h_{1, \text{exact} }$ & & $y_{1, \text{extr} }$ & & $y_{1, \text{guess} }$   \\
   \hline
   $3.5$           & & $10^{\,-7}$ &  & $0$             & &  $1.001$     & &  $1$    \\
   $3.0$           & & $10^{\,-7}$ &  & $0$             & &  $1.007$     & &  $1$    \\
   $K_0 (3) = 4\,\ln 2 = 2.772\cdots$
                   & & $10^{\,-5}$ &  & $0$             & & $1.05$       & &  $1$     \\  \\
     $K$           & & $h_{2, \text{extr} }$           &  & $h_{2, \text{exact} }$  & & $y_{2, \text{extr} }$ & & $y_{2, \text{guess} }$   \\
    \hline   
   $3.5$           & & $10^{\,-7}$ &  & $0$             & & $1.0007$      & &  $1$       \\
   $3.0$           & & $10^{\,-6}$            &  & $0$             & & $1.004$       & &  $1$       \\
   $K_0  (3)= 4\,\ln 2 = 2.772\cdots$
                   & & $10^{\,-5}$            &  & $0$             & & $1.04$        & &  $1$       \\     \\
    $K$            & & $\;\;\;\;\;\;g_{\text{extr} }$ &  & $g_{\text{exact} }$  &  & $\;\;y_{g, \text{extr} }$  & &  $y_{g, \text{guess} }$   \\
  \hline  
   $3.5$           & & $0.13836168$           &  & $-$          &  & $0.997$             & &        $1$     \\
   $3.0$           & & $0.11401535$           &  & $-$          &  & $0.992$             & &        $1$      \\
   $K_0 (3) = 4\,\ln 2 = 2.772\cdots$
                   & & $0.08169236$           &  & $-$          & & $1.53$               & &        $3 / 2$  
  \end{tabular}
  \end{ruledtabular}
  \end{table*}
  \begin{table*}    
  \caption{The  results of extrapolation for $h < 0$: the real and imaginary part of the closest zeros, 
           $h_{1,2}$, followed by the corresponding convergence exponents.
           The error bars are estimated to be of order of the last cited digit. The numerical part of errors is by at least 
           one order of magnitude smaller than the overall errors (including also the finite-size corrections).
           \label{tb:q3RehImhgExtrCijela1Ising}}
  \begin{ruledtabular}
  \begin{tabular}{cclclclcccc}
  $K$             & & $h_{1, \text{extr}  }$   &  & $h_{1, \text{exact} }$  & & $y_{1, \text{extr} }$\\ \hline
  $2.7$           & & $-0.04162$      &  & $-0.04169\cdots$& & $1.205$       \\
  $2.6$           & & $-0.1093$       &  & $-0.1099\cdots$ & & $0.685$       \\
  $2.5$           & & $-0.192$        &  & $-0.193\cdots$  & & $0.479$        \\
                  & &                 &  &                 & &            \\
    $K$           & & $h_{2, \text{extr} }$   &  & $h_{2, \text{exact} }$  & & $y_{2, \text{extr} }$\\ \hline
  $2.7$           & & $\;\;\;10^{-4}$ &  & $0$             & & $1.068$      \\
  $2.6$           & & $\;\;\;10^{-4}$ &  & $0$             & & $0.79$        \\
  $2.5$           & & $\;\;\;10^{-3}$ &  & $0$             & & $0.69$   
  \end{tabular}
  \end{ruledtabular}
  \end{table*}
  A qualitatively different transition, discussed in Sec. \ref{sec-IsingTransitions} appears for $h < 0$ 
  (Table \ref{tb:q3RehImhgExtrCijela1Ising}, open circles in Fig. \ref{fg:FazniDijagKh}). 
  The numerical precision of the results in Table \ref{tb:q3RehImhgExtrCijela1Ising} is much poorer than that in Table 
  \ref{tb:q3RehImhgExtrKMS}. 
  The main  reason for that could be attributed to the fact that the corresponding zeros are fewer and more spaced so that  
  even the closest two zeros are much more distant from the real axis, than they are for another line. 
  For this reason, we do not proceed here with the calculation of the density of zeros used for other cases. \\
  The numerical extrapolations  $h_{1, 2, \text{extr} }$ and $h_{1, 2, \text{exact}}$  presented in  Table 
  \ref{tb:q3RehImhgExtrCijela1Ising} yield the exact values  [given by (\ref{eq:ParcfPox}) and (\ref{eq:ParcfPoS})]  
  with numerical precision up to three digits or more.  The convergence exponents $y_{1, 2, \text{extr} }$  differ, 
  however, significantly from the scaling exponent  of the first-order phase transition,
  which should persist up to $h = - \infty$, as argued in Sec. \ref{sec-IsingTransitions}.

  \section{Conclusion} \label{sec-zklj}

  The phase diagram of the two- and three-state Potts model with infinite-range interactions in the external
  field was analyzed by studying the partition function zeros in the complex field plane. 

  In the three-state case, we derived the exact two-pa\-ra\-me\-ter expression for the free-energy density 
  after a twofold application of the Hubbard-Stratonovich transformation.
  By applying the saddle-point approximation in the two-pa\-ra\-me\-ter space, we reproduced well the known 
  analytical results in different regimes of the phase diagram with zero and positive external field. Calculations 
  could also be extended to the negative fields, where another, Ising-like, phase transition occurs. We derived
  an implicit analytical expression and performed exact numerical calculation for the critical line of this Ising-like 
  transition, showing that it is of first order, and becomes of second order only in the
  limit $h\rightarrow - \infty$, unlike the case of the three-dimensional Potts model with 
  short-range interactions, where the tricritical value of the field is finite \cite{BD2010}.

  The same procedure was efficient for the study of the partition function zeros in the complex field plane.
  We have shown, that in this model, with a rather complex phase diagram and with consequently more complicated loci 
  of zeros including multiple lines, it is still possible to identify different regimes of critical behavior from the
  size scaling properties of the few zeros closest to the real axis. 
  We successfully identified the regime of the first-order phase transitions, and well reproduced the critical exponents 
  belonging to the two second-order phase transitions, the tricritical point, and the Yang-Lee edge singularity.
  In the case of the tricritical point, an unusual feature is that the line of zeros reaches the real axis at the point with
  nonzero field (i.e. loci of zeros lie on the non unit circle in the complex activity plane).
  The Yang-Lee edge singularity could be observed for temperatures above the tricritical point temperature ($K < K_{tr}$),
  and was found to belong to the same universality class as that of the MF Ising model. 

  Due to the fact that the external field $h$ breaks the symmetry between Potts states of the $q = 3$ model in essentially 
  the same way as for all $q > 3$ models, one can expect that the general shape of the phase diagram presented in figure
  \ref{fg:FazniDijagKh} as well as a major part of the obtained results could be extended to models with $q > 3$. 

  It would also be interesting to perform a complementary study of a more general case of the Yang-Lee zeros for the
  Potts model with infinite range interactions with the power-law decay, where some previous work has been performed 
  \cite{GU91} for the Ising model by the finite-range scaling approach \cite{UG88}.

  \begin{acknowledgements}

  This work was supported by the Croatian Ministry of Science, Education and 
  Sports through grant No. 035-0000000-3187.

  \end{acknowledgements}



\end{document}